 \documentclass[12pt]{article}
 \usepackage{url}
  
\catcode`@=11
\def\secteqno{\@addtoreset{equation}{section}%
\def\theequation{\thesection.\arabic{equation}}}
\catcode`@=12 \secteqno
\newcommand{\be}{\begin{equation}}
\newcommand{\ee}{\end{equation}}
\newcommand{\bea}{\begin{eqnarray}}
\newcommand{\eea}{\end{eqnarray}}
\newcommand{\bref}[1]{(\ref{#1})}
\newcommand{\nn}{\nonumber}

\def\repnum{galileiallcohomology.tex (\today)}
\newcommand{\bi}{\begin{enumerate}}
\newcommand{\ei}{\end{enumerate}}

\newcommand{\h}{\eta}

\def\6{\partial}
\def\7{\tilde}
\def\8{\hat}

\def\={{\;=\;}}\def\+{{\;+\;}}

\begin{document}
\thispagestyle{empty} \hfill\today

\hfill UB-ECM-PF/08-15

\vskip 20mm
\begin{center}
{\Large\bf A note on the Chevalley-Eilenberg Cohomology for the
Galilei and Poincar{\' e} Algebras} \vskip 10mm
{\large ~ Sotirios Bonanos and Joaquim
Gomis} \vskip 6mm
\medskip

Institute of Nuclear Physics, NCSR Demokritos,\\
15310 Aghia Paraskevi, Attiki, Greece\\ and \\
Departament d'Estructura i Constituents de la
Mat\`eria and ICCUB, Universitat de Barcelona, Diagonal 647, 08028 Barcelona\\
\vskip 6mm
 {\small e-mail:\
 {sbonano@inp.demokritos.gr\\gomis@ecm.ub.es} }\\

\medskip

\end{center}
\vskip 40mm
\begin{abstract}
We construct in a systematic way  the complete Chevalley-Eilenberg
cohomology at form degree two, three and four for the Galilei and Poincar{\' e} groups.
The corresponding
non-trivial forms belong to certain representations of the spatial
rotation (Lorentz) group. In the case of two forms they give all possible
central and non-central extensions of the Galilei group (and all non-central extensions of the
 Poincar{\' e} group). The
procedure developed in this paper can be applied to any space-time
symmetry group.

\end{abstract}

\vskip 4mm

\setcounter{page}{1}
\parskip=7pt
\newpage
\section{Introduction and Results}
 Physical theories are characterized by their symmetries. The basic
symmetry is that of the underlying space: the transformations that
are implied by the homogeneity and isotropy of space. Physical
theories respect these kinematical symmetries and exhibit additional
dynamical symmetries.  Among the possible kinematical groups, which
have been classified in
 \cite{Bacry:1968zf}, the Poincar{\' e} and Galilei
groups are  the better known ones. Here we will study the central
and non-central  extensions of these two groups,  obtaining extended
symmetries which can be used to construct new physical theories  or
to interpret group theoretically certain symmetry properties of
known theories.

 It is  well known that the Galilei group in d+1 space-time
dimensions admits a central extension \cite{Bargmann:1954gh}, while
in the special case of 2+1 dimensions, it admits an exotic, two-fold
central extension \cite{LL, BGO, BGGK}. This exotic
Galilei group has attracted attention in the context and
non-commutative geometry and in  condensed matter physics
\cite{LSZ,DH1,JackN,HP1}, see also \cite{Horv} for further references.
 In the particular case of $1+1$
dimensions the Galilei algebra has two central extensions and the
Poincar{\' e} algebra one\footnote{These $1+1$ extended algebras
have been used  to study several problems of gravity and Moyal
quantization, see for example
\cite{Aldaya:1985va,olmo1,Cangemi:1992bj}.} \cite{levyleblond69}.
The Poincar{\' e} group has no central extensions in d+1 ($d>1$)
space-time dimensions; however it has a non-central antisymmetric
tensor extension,  which has an important physical application: it
describes the motion of a relativistic particle in a constant
electromagnetic field \cite{Bacry:1970ye,
Schrader:1972zd,Beckers:1983gp}, see also \cite{Soroka:2004fj}.

 The extensions of a group are  related to a non-trivial
Chevalley-Eilenberg (CE)  {cohomology} of the corresponding
unextended Lie algebras,  for a review see  \cite{azcarragabook}. In
this paper we compute in a systematic, almost algorithmic,  way the
most general CE cohomology groups at form degree  two, three and
four. As we will see the non-trivial forms belong to different
representations of the subgroup of spatial rotations (Lorentz) of
the Galilei (Poincar{\' e}) group.

In all space-time algebras we can separate the generators into
 ``translations" and  ``rotations". The  ``rotations" generators are a
subgroup  of the automorphism group and  constitute a normal
subgroup of  every extended  group. In the case of the Galilei group
we consider as ``translations" the space and time translations and
the galilean boosts, while the  ``rotations" are the ordinary
spatial rotations. For the Poincar{\' e} group  ``translations" are
the space-time translations and the  ``rotations" are the Lorentz
transformations.

The first extension, central and non-central, of the  unextended
Galilei algebra is obtained by calculating  all possible non-trivial
closed 2-forms of the subgroup of  ``translations". The forms are
closed with respect the exterior differential operator $d$. These
forms can be grouped into different representations of the rotation
group:  a vector, an antisymmetric tensor and a symmetric tensor.
(In the case of Poincar{\' e} there is  only  an antisymmetric tensor
representation.) The trace of the symmetric tensor corresponds to
the well-known central extension of the Galilei group
\cite{Bargmann:1954gh}. In the case of 2+1 dimensions the
antisymmetric tensor becomes a pseudoscalar representation which
gives rise to the second central extension.

The complete extended algebra is constructed from the original
algebra and the extensions by incorporating their transformation
properties  under rotations, which is equivalent to replacing the
exterior
 differential operator $d$ by the corresponding  ``covariant"
 operator $d+M \wedge$, $M$ being the zero-curvature connection
associated to the  ``rotations": $dM+M\wedge M=0$.

As we will see, once we have an extended algebra, we can further
extend it by applying the previous procedure to the new set of
translations that now includes the extended generators. In this way
we obtain new extensions with some generators belonging to higher
dimensional representations of the  corresponding  ``rotations"
group. The physical significance of these higher dimensional
extensions should be clarified.
At this level the generators of the first and second level
extensions couple non-trivially to each other. This procedure for the
case of the Galilei and Poincar{\' e} groups does not end.

We have also studied the first level CE cohomology at degrees three and four. As in the
case of 2-forms,  the non-trivial 3- or 4-forms obtained are grouped into
representations of the  ``rotations" group. The non-invariant
2-form ``potentials"  associated to these closed 3-forms could be used to construct
non-relativistic string theories whose Lagrangian will be invariant
up to a total derivative under the Galilei (Poincar{\' e}) group.

For the Galilei group these string theories will be different from
the non-relativistic string theory introduced in
\cite{Gomis:2000bd}, \cite{Danielsson:2000gi}, \cite{Gomis:2005pg}
because the invariance group of this string theory is the  Galilean
string group which is obtained by a stringy contraction of the
Poincar{\' e} group \cite{Brugues:2004an} and not by the ordinary
contraction that leads to the ordinary Galilei algebra. In the case
of the Poincar{\' e} group the forms can be used to construct
Wess-Zumino (WZ) terms for relativistic string theories coupled to
external backgrounds. Finally, we note that the non-trivial closed
4-forms obtained here could be used to construct WZ   actions  of
non-relativistic and relativistic branes.

The procedure to compute CE cohomolgy groups
 employed in this note is described in the Appendix and can be
applied to any space-time symmetry group.  The results are given for the case of 3+1
dimensions, but   this procedure, mutatis mutandis, can be used   in any number of space dimensions.
The application to super space-time groups  with odd generators will be done in a
forthcoming paper.

The organization of the paper is as follows. In section 2 we give
our notation and conventions  and study the first level extension of
the Galilei group. In section 3 we analyze the second level
extension.  In section 4 we  give the results of the first level  CE cohomology
at form-degree three and four, while in section 5 we give the
corresponding results for the Poincar{\' e} group.
 Finally, in an Appendix,
we describe the  ``algorithm" used for obtaining these extensions. The
calculations  make use of the first author's Mathematica  package EDC
(Exterior Differential Calculus) \cite{bonanos}.

\section{Galilei group in 3+1 dimensions}
The generators of the unextended Galilei algebra  are the
hamiltonian, $H$, the spatial translations $P_{a}$, the boots
$K_{a}$, and the rotations $M_{ab}$. The algebra is given
by\footnote{ Although the algebra is real and the imaginary units
can be made to disappear by replacing all generators $G$ by $i G'$,
we prefer to leave the $i$'s in the equations because then we can
interpret the generators as Hermitian operators.}
\bea
\left[M_{ab},~M_{cd}\right]&=&-i~\h_{b[c}~M_{ad]}+i~\h_{a[c}~M_{bd]},
\nn\\
\left[K_{a},~M_{cd}\right]&=&-i~\h_{a[c}~K_{d]},
\nn\\
\left[P_a,~M_{bc}\right]&=&-i~\h_{a[b}~P_{c]},
\nn\\
\left[H,~M_{ab}\right]&=&0,
\nn\\
\left[H,~K_{a}\right]&=&+i~P_{a}, \\
\left[H,~P_{a}\right]&=&0,
\nn\\
\left[K_{a},~K_{b}\right]&=&0,
\nn\\
\left[P_a,~K_{b}\right]&=&{0},
\nn\\
\left[P_a,~P_{b}\right]&=&0, \nn \label{NHm} \eea where $\h_{ab}$ is the
euclidean metric of 3-dimensional space, $a,b=1,\,2,\,3$.

In order to construct the extensions it is very useful  to introduce
the left invariant Maurer-Cartan (MC) form, defined by \be
\label{omega}\Omega=-i g^{-1}dg,\ee where $g$ represents a general
element of the Galilei group. The MC form satisfies the
Maurer-Cartan equation \be
d\Omega+i~\Omega\wedge\Omega=0.\label{MCeq}\ee
In components the MC 1-form is written, for a generic Lie algebra,
as \be \label{MC1}\Omega= X_A~ \cal{X}^A,\label{compMCeq} \ee where $X_A$ are the
generators of the Lie algebra satisfying  \be [X_B,~X_C]=i~{f^A}_{BC}~X_A \ee
and $\cal{X}^A$ are corresponding 1-forms. (Throughout this paper we use the same capital letters in plain and calligraphic font  to denote generators and associated 1-forms). The MC equation  \bref{MCeq}  implies
that the 1-forms  $\cal{X}^A$ satisfy
\be d{\cal X}^A=\frac12{f^A}_{BC} {\cal X}^B\wedge {\cal X}^C. \ee

For the Galilei case, the MC 1-form \bref{MC1} becomes \be \Omega=
H~{\cal H}+P_a~{\cal P}^a+K_a ~{\cal K}^a+\frac12 M_{ab}~{\cal
M}^{ab},\ee while the MC equation  \bref{MCeq} in components is
\bea
d{\cal H}&=&0,
\nn\\
d{\cal P}^a&+&{\cal P}^c \wedge{{\cal M}_c}^a-{\cal H} \wedge{\cal K}^a=0,
\nn\\
d{\cal K}^a&+&{\cal K}^c \wedge{{\cal M}_c}^a\;=0,
\nn\\
d{\cal M}^{ab}&+&{\cal M}^{ac} \wedge{{\cal M}_c}^b\;=0.
\label{MCNH} \eea

 In order to start our procedure for the cohomological analysis we
 need to define the   ``translations" generators, which we take to be the space and time
translations and the boosts, i.e.,  $H, P_a, K_a$. The MC equations
for these generators are  obtained by putting ${\cal
M}^{ab}\rightarrow 0$ in \bref{MCNH}.

The most general closed invariant 2-form which cannot be written as
the differential of an invariant 1-form is then found to be

\be\label{general2form} \Omega_2= f_a\,{\cal H}\,\wedge{\cal P}^a+
f_{\left[ab \right]}\,{\cal K}^a\wedge{\cal K}^b+ f_{(ab)}\,{\cal
K}^a\wedge {\cal P}^{b} \ee where the constant parameters are a
vector $f_a$, a second rank antisymmetric tensor $f_{\left[ab
\right]}$ and a second rank symmetric tensor $f_{(ab)}$\footnote{The
calculation proceeds in several distinct steps, as outlined in the
Appendix.}. Therefore we find that
 the non-trivial 2-forms belong to a vector, a symmetric and an
antisymmetric tensor representation of the rotation group. The
1-form  ``potentials" associated to these 2-forms are denoted by \be
{\cal Z}^a, ~{\cal Z}^{\left[ab \right]},~{\cal Z}^{(ab)}\ee and
satisfy he MC equations \bea d{\cal Z}^a&=&{\cal H}\wedge{\cal
P}^a\nn\\
d{\cal Z}^{\left[ab \right]}&=&{\cal K}^a\wedge{\cal K}^b \label{dZeqsM0}\\
d{\cal Z}^{(ab)}&=&{\cal K}^a\wedge {\cal P}^{b}+{\cal K}^b\wedge
{\cal P}^{a}.\nn \eea

From these expressions we can
obtain the algebra of  the corresponding generators, denoted by
\be Z_a, ~Z_{\left[ab \right]} , ~Z_{(ab)}.\ee We find
 \bea \left[K_{a},~K_{b}\right]&=& +i~Z_{\left[ab
\right]} \nn\\ \left[P_a,~K_{b}\right]&=& -i~Z_{(ab)},\\
 \left[H,~P_{a}\right]&=&+i~ Z_a.\nn \label{NHm1} \eea
 The extension $Z_a$ has been used to describe the motion of a particle in
a constant field,  $Z_{\left[ab \right]}$  to study the motion of a
particle in a non-commutative space, while $Z_{(ab)}$ appears in
the description of  a particle with mass anisotropy. The ordinary
central extension of the Galilei group corresponds to the trace part
of $Z_{(ab)}$. In 2 +1 dimensions, the second (exotic) central
extension  corresponds to the single independent component of the
antisymmetric tensor $Z_{\left[ab\right]}$,
 while in 1+1 dimensions the single-component vector extension
$Z_a$  gives the extra central charge of Galilei
\cite{levyleblond69}.

With the rotations included, the extended set of MC 1-forms satisfies the equations
 \bea d{\cal H}&=&0,
\nn\\
d{\cal P}^a&=&-{\cal P}^c \wedge{{\cal M}_c}^a+{\cal H} \wedge{\cal K}^a,
\nn\\
d{\cal K}^a&=&-{\cal K}^c \wedge{{\cal M}_c}^a,
\nn\\
d{\cal M}^{ab}&=&-{\cal M}^{ac} \wedge{{\cal M}_c}^b,\label{extMCNH}\\
d{\cal Z}^a&=&-{\cal Z}^c \wedge{{\cal M}_c}^a+{\cal H}\wedge{\cal P}^a \nn\\
d{\cal Z}^{\left[ab \right]}&=&-{\cal Z}^{\left[ac \right]} \wedge {{\cal M}_c}^b-
{{\cal M}^a}_c \wedge{\cal Z}^{\left[cb \right]} +{\cal K}^a \wedge{\cal K}^b, \nn\\
d{\cal Z}^{(ab)}&=&-{\cal Z}^{(ac)} \wedge {{\cal M}_c}^b- {{\cal M}^a}_c \wedge{\cal Z}^{(cb)}+
{\cal K}^a\wedge {\cal P}^{b}+{\cal K}^b\wedge{\cal P}^{a}.\nn
\eea

\subsection{Explicit parametrization}
Here we will obtain an explicit parametrization for all MC 1-forms in terms of
differentials of functions, so that equations \bref {extMCNH} are satisfied.
 We first obtain expressions for the unextended MC 1-forms without rotations.
Locally, we can parametrize a  general element of the unextended group by
 \be\label{parametrization0}
g=e^{iHx^0}e^{iP_ax^a}e^{iK_{a}v^a}. \ee The MC 1-form \bref{omega}
can be computed directly from its definition using the
Baker-Campbell-Hausdorff formula. The components of the MC 1-form
\bref{compMCeq} can be computed explicitly using  Mathematica and EDC
\cite{bonanos}, if we substitute  a matrix representation for the
generators, for example
 the adjoint representation. The result is
\bea {\cal H}&=&dx^0,\nn\\
 {\cal P}^a&=&dx^a-{v^a} dx^0,\label{explicit0}\\
 {\cal K}^a&=&dv^a.\nn
\eea

For the extended group, instead of following the same procedure
starting with the general element  \be\label{parametrization}
g=e^{iHx^0}e^{iP_ax^a}e^{iK_{a}v^a}
e^{iZ_{a}k^a}e^{iZ_{(ab)}k^{(ab)}}e^{iZ_{\left[ab
\right]}k^{\left[ab \right]}}, \ee it is much easier to use the
known parametrization for the 1-forms ${\cal H}, \; {\cal P}^a,\;
{\cal K}^a$ on the  right hand side  of \bref{dZeqsM0} and integrate.
The result is that the new MC 1-forms, modulo exact 1-forms, can be
written
\bea
{\cal Z}^a&=&dk^a-x^a\; dx^0,\nn\\
{\cal Z}^{\left[ab \right]}&=& dk^{\left[ab \right]}+\frac12(v^a\;dv^b-v^b\;dv^a), \label{explicit1}\\
{\cal
Z}^{(ab)}&=&dk^{(ab)}+v^a\;d(x^b-v^b\;x^0)+v^b\;d(x^a-v^a\;x^0),
\nn\eea where $k^a,\;k^{\left[ab \right]},\;k^{(ab)}$ are new functions --  the group
parameters associated to the new generators in \bref{parametrization}.
These 1-forms can be used to construct non-relativistic particle
Lagrangians. The study of these Lagrangians and their physical
implications will be considered elsewhere. Here let us notice that
the familiar free non-relativistic massive particle is obtained by
taking the trace of ${\cal Z}^{(ab)}$ and coincides with the one
obtained with the method of non-linear realizations \cite{Coleman},
see for example \cite{Gauntlett:1990nk}.

If we want to have the explicit expressions of the MC 1-forms when
we include rotations, the   right hand side   of all vector and
tensor expressions given above must be
 multiplied by an appropriate    rotation matrix for each index:
\be {\cal P}^a={{U^{-1}}^a}_b\;(dx^b+\cdots),\hspace{30 pt} {\cal
Z}^{ab}= {{U^{-1}}^a}_p\;{{U^{-1}}^b}_q\;(dk^{pq}+\cdots),\ee where
the rotation MC 1-forms   have the explicit representation
 ${{\cal M}^{a}}_{b}= {{U^{-1}}^a}_c\; d[{U^c}_b]$.

\section{Second Level Extensions}

One can obtain further extensions of the Galilei group corresponding to
non-trivial 2-forms in higher dimensional representations of the
rotation group. In order to find them we follow the same procedure as in the last
section, this time taking as  ``translations" the nineteen
 1-forms \be {\cal H},~{\cal
P}^a,~{\cal K}^a, ~{\cal Z}^a,~ {\cal Z}^{\left[ab \right]},~{\cal
Z}^{(ab)}.\ee

We obtain 33 new
non-trivial closed 2-forms which are the components of the following tensors:
 \bea
{\cal H} \wedge{\cal Z}^a, \hspace{55 pt} \nn \\
{\cal H} \wedge{\cal Z}^{(ab)}+{\cal K}^a\wedge{\cal Z}^b+{\cal K}^b\wedge{\cal Z}^a, \nn \\
{\cal P}^a \wedge{\cal P}^b+{\cal K}^a\wedge{\cal Z}^b-{\cal K}^b\wedge{\cal Z}^a, \nn \\
{\cal H} \wedge{\cal Z}^{\left[ab \right]}+{\cal K}^a \wedge{\cal P}^b-{\cal K}^b \wedge{\cal P}^a,  \\
{\cal K}^a \wedge{\cal Z}^{(bc)}+{\cal K}^b \wedge{\cal
Z}^{(ca)}+{\cal K}^c \wedge{\cal Z}^{(ab)}, \nn \\
2~{\cal Z}^{\left[ab \right]}\wedge
{\cal K}^c-{\cal Z}^{\left[bc \right]}\wedge {\cal K}^a-{\cal Z}^{\left[ca \right]}\wedge {\cal K}^b,\nn
 \eea the latter being a tensor antisymmetric in the first two indices ($ab$) whose totally antisymmetric part  $\epsilon_{abc}~ {\cal Z}^{\left[ab \right]}\wedge {\cal K}^c$ vanishes.

If we use the kernel symbol ${\cal Y}$ for the 1-form tensor
 ``potentials" associated to these 2-forms and include the effect of rotations,
we find that they satisfy the equations
\bea
d{\cal Y}^a&=&-{\cal Y}^c \wedge{{\cal M}_c}^a+{\cal H} \wedge{\cal Z}^a,\nn \\
d{\cal Y}^{(ab)}&=&-{\cal Y}^{(ac)} \wedge {{\cal M}_c}^b- {{\cal M}^a}_c \wedge{\cal Y}^{(cb)}+{\cal H} \wedge{\cal Z}^{(ab)}+{\cal K}^a\wedge{\cal Z}^b+{\cal K}^b\wedge{\cal Z}^a, \nn \\
d{{\cal Y}_1}^{\left[ab \right]}&=&-{{\cal Y}_1}^{\left[ac \right]} \wedge {{\cal M}_c}^b- {{\cal M}^a}_c \wedge{{\cal Y}_1}^{\left[cb \right]}+{\cal P}^a \wedge{\cal P}^b+{\cal K}^a\wedge{\cal Z}^b-{\cal K}^b\wedge{\cal Z}^a, \nn \\
d{{\cal Y}_2}^{\left[ab \right]}&=&-{{\cal Y}_2}^{\left[ac \right]} \wedge {{\cal M}_c}^b- {{\cal M}^a}_c \wedge{{\cal Y}_2}^{\left[cb \right]}+{\cal H} \wedge{\cal Z}^{\left[ab \right]}+{\cal K}^a \wedge{\cal P}^b-{\cal K}^b \wedge{\cal P}^a, \nn \\
d{\cal Y}^{(abc)}&=&-{\cal Y}^{(abs)} \wedge {{\cal M}_s}^c-{\cal Y}^{(bcs)} \wedge {{\cal M}_s}^a-{\cal Y}^{(cas)} \wedge {{\cal M}_s}^b \label{extYeqs} \\ &\;&+{\cal K}^a \wedge{\cal Z}^{(bc)}+{\cal K}^b \wedge{\cal
Z}^{(ca)}+{\cal K}^c \wedge{\cal Z}^{(ab)},\nn \\
d{{\cal Y}_3}^{\left[ab \right]c}&=&- {{\cal M}^a}_s \wedge{{\cal Y}_3}^{\left[sb \right]c} - {{\cal M}^b}_s \wedge{{\cal Y}_3}^{\left[as \right]c} - {{\cal M}^c}_s \wedge{{\cal Y}_3}^{\left[ab \right]s}\nn \\ &\;&
+2~{\cal Z}^{\left[ab \right]}\wedge
{\cal K}^c-{\cal Z}^{\left[bc \right]}\wedge {\cal K}^a-{\cal Z}^{\left[ca \right]}\wedge {\cal K}^b,\nn
\eea
where the symmetries of the  ${\cal Y}$ tensors are as indicated by round or square brackets, while the totally antisymmetric part of ${{\cal Y}_3}^{\left[ab \right]c}$ vanishes. The corresponding new algebra generators  will be written \be Y_a, ~Y_{(ab)},~ {Y^1}_{[ab]}, ~ {Y^2}_{[ab]}, ~Y_{(abc)} ,~  {Y^3}_{\left[ab \right]c}. \ee

We observe that, at this level the generators of the extensions
couple non-trivially to each other, for example \be
\left[K_a,~Z_{(bc)} \right]=i~Y_{(abc)}, \ee while the $P_a$ no
longer commute:  \be \left[P_a,~P_b \right]=i~{Y^1}_{[ab]}.\ee The
generator  ${Y^1}_{[ab]}$ can be used to describe the motion of
non-relativistic particles in a constant magnetic background.

We could complete the algebra by introducing the transformations of the
new generators under rotations. For the case of the Galilei group
this extension procedure does not  end. The physical interpretation
of these new {\cal Y} terms remains to be understood.

\section{CE cohomology at degrees three and four}

In this section we study the most general non-trivial
closed 3-forms and 4-forms.
Our procedure gives 18 closed non-trivial 3-forms, which, as in the case of 2-forms studied above,
can be grouped under different representations of the rotation group, resulting in:
\begin{itemize}
\item {a symmetric tensor ${\cal H} \wedge({\cal K}^a \wedge {\cal P}^b+{\cal K}^b \wedge {\cal P}^a)$}
\item {an antisymmetric tensor ${\cal H} \wedge{\cal P}^a \wedge {\cal P}^b$}
\item {a pseudoscalar $\epsilon_{abc}\;{\cal K}^a \wedge{\cal K}^b \wedge {\cal K}^c$}
\item {and the following 3rd rank tensor, antisymmetric in its first 2 indices $\left[ab\right]$,
 whose totally antisymmetric part vanishes
 \be2~{\cal K}^a \wedge{\cal K}^b \wedge {\cal P}^c-{\cal K}^b \wedge{\cal K}^c
 \wedge {\cal P}^a-{\cal K}^c \wedge{\cal K}^a \wedge {\cal P}^b. \ee}
\end{itemize}
These closed 3-forms imply the existence of corresponding 2-form
tensor potentials\footnote{Closed differential systems of tensor valued differential forms of any order are known as "Free Differential Algebras" (FDAs), see  \cite{AuriaFre}  \cite{Castellani}.}. In terms of the explicit parametrization of group
elements introduced in \bref{explicit0}, \bref{explicit1}, we can
write these 2-form potentials  (modulo exact of 2-forms) as
 \bea
 -(v^a\;dx^0\wedge dx^b +v^b\;dx^0\wedge dx^a), \hspace{45 pt} \nn \\
 -\frac12(x^a\;dx^0\wedge dx^b -x^b\;dx^0\wedge dx^a), \hspace{40 pt} \nn\\
\epsilon_{abc}\;v^a\;dv^b \wedge \;dv^c, \hspace{80 pt} \\
vvx(a,b,c)- \frac12 vvx(b,c,a)- \frac12 vvx(c,a,b), \hspace{30 pt}\nn
\eea where $vvx(a,b,c)$ stands for $(v^a\;dv^b-v^b\;dv^a)\wedge(dx^c -v^c\;dx^0)$.

These 2-forms could be used to construct non-relativistic string
theories whose Lagrangian is invariant up to an exact form under the
Galilei group. As mentioned in the Introduction, these string
theories will be different from the non-relativistic string theory
introduced in \cite{Gomis:2000bd}, \cite{Danielsson:2000gi},
\cite{Gomis:2005pg}.

Finally, applying our procedure at degree four we obtain 19 non-trivial
closed 4-forms which are the
components of the following tensors: \bea
{\cal H} \wedge{\cal K}^a \wedge {\cal P}^b \wedge{\cal P}^c, \hspace{55 pt} \nn \\
{\cal H} \wedge \epsilon_{abc} \;{\cal P}^a \wedge {\cal P}^b \wedge{\cal P}^c, \hspace{55 pt} \nn \\
{\cal P}^d  \wedge \epsilon_{abc} \;{\cal K}^a \wedge {\cal K}^b \wedge{\cal K}^c, \hspace{55 pt}  \\
 {\cal P}^a \wedge {\cal P}^b \wedge {\cal K}^c \wedge {\cal K}^d + {\cal K}^a \wedge {\cal K}^b \wedge {\cal P}^c \wedge {\cal P}^d.  \nn
\eea
The corresponding non-trivial forms  could be used for constructing WZ
terms of non-relativistic branes. In higher dimensions, the terms containing $\epsilon_{abc}$ would give totally antisymmetric third rank tensors.

\section{Poincar{\' e} Extensions}
It is known that the Poincar{\' e} group admits no
 central extensions in $d>1$ dimensions. However it has an
antisymmetric tensor non-central extension
\cite{Bacry:1970ye,Schrader:1972zd,Beckers:1983gp,Soroka:2004fj}.
 Here we apply our procedure to the Poincar{\' e} group and obtain
 all non-central extensions. We will use the same notation as in the previous sections, but the tensor
 indices will  now take the values $(0,\;1,\;2,\;3)$ while  $\h_{ab}$ will denote
 the Minkowski metric.  The generators now satisfy the equations\bea
\left[M_{ab},~M_{cd}\right]&=&-i~\h_{b[c}~M_{ad]}+i~\h_{a[c}~M_{bd]},\nn\\
\left[P_a,~M_{bc}\right]&=&-i~\h_{a[b}~P_{c]}\label{Poincare0} \eea
and the corresponding MC 1-forms satisfy \bea
d{\cal P}^a&=&-{{\cal M}^a}_c\wedge{\cal P}^c,\nn\\
d{\cal M}^{ab}&=&-{\cal M}^{ac} \wedge{{\cal M}_c}^b.\label{Poincare1}
\eea
Freezing the Lorentz freedom (${{\cal M}^a}_b\rightarrow 0$), the first-level
 extension results in the  antisymmetric tensor 2-form
 ${\cal P}^a\wedge{\cal P}^b$.
 Using the same notation for the corresponding potential 1-form
 (${\cal Z}^{\left[ab \right]}$) and reintroducing, as before,  the Lorentz freedom, we arrive at
\be d{\cal Z}^{\left[ab \right]}=- {{\cal M}^a}_c \wedge{\cal
Z}^{\left[cb \right]} - {{\cal M}^b}_c \wedge{\cal Z}^{\left[ac
\right]}+{\cal P}^a \wedge{\cal P}^b, \ee which implies that the
generators of translations no longer commute: \be \left[P_a,~P_b
\right]= i~{Z}_{[ab]}.\ee As there are no other extensions, the well
known result that there are no central extensions for the Poincar{\'
e} group follows.  This antisymmetric generator ${Z}_{[ab]}$ can be
used to describe the motion of a relativistic particle in a constant
electromagnetic field.

The calculation of the second level extensions
results in 20 closed non-trivial 2-forms which
can be written as the components  the tensor\footnote{This tensor is antisymmetric
in $\left[bc\right]$ and its totally antisymmetric part
vanishes. This leads to 4 identities, $\epsilon_{abcd} {{\cal P}^b}
\wedge{\cal Z}^{\left[cd \right]}=0$, leaving 4x6-4=20 independent
components.}

\be  2~{{\cal P}^a} \wedge{\cal Z}^{\left[bc \right]}  - {{\cal P}^b}
\wedge{\cal Z}^{\left[ca \right]}-{{\cal P}^c}\wedge{\cal Z}^{\left[ab \right]}. \ee
Again, introducing the
second-level potential 1-form ${\cal Y}^{a\left[bc\right]}$ with the same
symmetries as the above 2-form tensor and unfreezing the Lorentz
freedom, we find \bea d{\cal Y}^{a\left[bc\right]}&=&-{{\cal M}^a}_s \wedge{\cal
Y}^{s\left[bc\right]} -{{\cal M}^b}_s \wedge{\cal Y}^{a\left[sc\right]}-{{\cal M}^c}_s
\wedge{\cal Y}^{a\left[bs\right]} \nn \\& & +~2~{{\cal P}^a} \wedge{\cal Z}^{\left[bc \right]}
 - {{\cal P}^b}
\wedge{\cal Z}^{\left[ca \right]}-{{\cal P}^c}\wedge{\cal Z}^{\left[ab \right]}. \eea
 As in the Galilei case, the generators of first and second level extensions
 now  couple non-trivially  to each other \be
 \left[P_a,~Z_{\left[bc \right]}  \right]=2~i~Y_{a\left[bc\right]}-i~
 Y_{b\left[ca\right]}-i~Y_{c\left[ab\right]}.\ee

Finally, the CE cohomology  at higher degrees gives simple non-trivial results. For
 degree 3 there are 4 closed  3-forms which are the
components of the pseudovector \be \epsilon_{abcd}\;{{\cal P}^b}
\wedge{{\cal P}^c} \wedge{{\cal P}^d}, \ee while at degree 4 there is a single pseudoscalar
4-form:  $ \epsilon_{abcd}\;{{\cal P}^a} \wedge{{\cal P}^b}\wedge{{\cal P}^c} \wedge{{\cal P}^d} $.

\section*{Acknowledgments}
We acknowledge Roberto Casalbuoni, Gary Gibbons, Kiyoshi Kamimura,
Giorgo Longhi and Chris Pope for useful discussions about cohomology
and symbolic computation.  We also thank the referees for their
comments and several new references. This work has been supported by
MCYT FPA 2007-66665, CIRIT GC 2005SGR-00564, Spanish
Consolider-Ingenio 2010 Programme CPAN (CSD2007-00042).

\section*{Appendix}
The calculations in this note are done using the Mathematica package
EDC (Exterior Differential Calculus) \cite{bonanos}
 and are contained in the Mathematica notebooks
GalileiSequentialExtensions.nb and PoincareSequentialExtensions.nb,
which are available from the authors. The calculations can be broken
down into the following five distinct steps (results given for the
Galilei case):

{\it Step 1: }
Set ${\cal M}^{ab}=0$ and construct all possible 2-forms from the set of ``translations" $\{{\cal H},~{\cal K}^a,~{\cal P}^a\}$, obtaining a set of twenty-one 2-forms. Find those  combinations of these 2-forms that are closed and not trivial (cannot be written as d[1-form]). This leads to the following twelve 2-forms
\bea
\{{\cal H}\wedge {\cal P}^1, ~{\cal H}\wedge {\cal P}^2, ~{\cal  H}\wedge {\cal P}^3, ~{\cal K}^1\wedge {\cal
          K}^2,~ {\cal K}^1\wedge {\cal K}^3, ~{\cal K}^2\wedge {\cal
      K}^3, ~{\cal K}^1\wedge {\cal P}^1, ~{\cal K}^2\wedge {\cal
      P}^2, \nn \\ {\cal  K}^1\wedge {\cal P}^2 + {\cal K}^2\wedge {\cal P}^1,~ {\cal
            K}^1\wedge {\cal P}^3 + {\cal K}^3\wedge {\cal P}^1, ~{\cal \
K}^2\wedge {\cal P}^3 + {\cal K}^3\wedge {\cal P}^2, ~{\cal K}^3\wedge {\cal   P}^3\} \nn
 \eea

{\it Step 2: }
Group these twelve 2-forms as the components of  a vector ${\cal H}\wedge {\cal P}^a$, an antisymmetric tensor  ${\cal K}^a \wedge {\cal K}^b$, and a symmetric tensor ${\cal K}^a \wedge {\cal P}^b+ {\cal K}^b \wedge {\cal P}^a$. This step is the only non-algorithmic part of the procedure and can be non-trivial.

{\it Step 3: }
Introduce the rotations and verify that these 2-forms remain ``closed" if we replace $d$ by the ``covariant" exterior derivative $d+{\cal M}\wedge $.

{\it Step 4: }
Now, if a tensor valued 2-form ${\cal F}^{a}$ satisfies the equation\footnote{The equations are written when ${\cal F}^{a}$ is a vector. For higher rank tensors additional ${\cal M}\wedge$ terms are needed -- see equations \bref{extMCNH} and \bref{extYeqs}.}  $d{\cal F}^{a}+{{\cal M}^a}_b \wedge {\cal F}^b=0$,  then there exists a corresponding tensor valued ``potential" 1-form ${\cal Z}^a$ satisfying
 \be d{\cal Z}^{a}+{{\cal M}^a}_b \wedge {\cal Z}^b={\cal F}^a. \ee
Introduce notation for the components of  these 1-form potentials and write the component form of these equations. Verify consistency.

{\it Step 5: }
Combine with the original algebra to obtain the extended algebra of MC 1-forms given in \bref{extMCNH}. Translate this algebra to the algebra of commutators of the corresponding generators.

The entire procedure of  5 steps is then repeated for the extended
set of 2-forms (second level extension). The space of 2-forms now consists
of 171 elements, of which 33 are found to be closed, non-trivial and
not included in the first extension. They are the components of the
tensors given in section 3.

Finally, the first two steps for the first-level 3-form and 4-form cohomology are also given. We find 18 closed non-trivial 3-forms and 19  closed non-trivial 4-forms which can be grouped as the components of the tensors given in section 4. Steps 3 and 4 (requiring now 2- or 3-form potentials) are straightforward.

\end{document}